\documentclass[epj]{svjour}
\usepackage{graphics}
\usepackage{epstopdf}
\usepackage{xcolor} 
\usepackage{cite}  
\begin{document}
\title{A Case Study of the May 30th, 2017 Italian Fireball}
\author{
Albino Carbognani\inst{1, 2}
D. Barghini \inst{2, 3}
D. Gardiol \inst{2}
M. Di Martino \inst{2}
G. B. Valsecchi\inst{4, 5}
P. Trivero \inst{6}
A. Buzzoni \inst{1}
S. Rasetti \inst{2}
D. Selvestrel \inst{7}
C. Knapic \inst{8}
E. Londero \inst{8}
S. Zorba \inst{8}
C. A. Volpicelli \inst{2}
M. Di Carlo \inst{9}
J. Vaubaillon \inst{10}
C. Marmo \inst{11}
F. Colas \inst{10}
D. Valeri \inst{12, 13}
F. Zanotti \inst{13}
M. Morini \inst{13}
P. Demaria \inst{13}
B. Zanda \inst{14}
S. Bouley \inst{11}
P. Vernazza \inst{15}
J. Gattacceca \inst{16}
J.-L. Rault \inst{17}
L. Maquet \inst{10}
M. Birlan \inst{10}
}                     
\offprints{albino.carbognani@inaf.it}          
\institute{INAF-Osservatorio di Astrofisica e Scienza dello Spazio,  Via Piero Gobetti, 93/3, 40129 Bologna, Italy
\and INAF-Osservatorio Astrofisico di Torino, Via Osservatorio 20, 10025 Pino Torinese (TO), Italy
\and Dipartimento di Fisica, Univ. degli studi di Torino, Italy
\and IAPS-INAF, via Fosso del Cavaliere 100, 00133 Roma, Italy
\and IFAC-CNR, via Madonna del Piano 10, 50019 Sesto Fiorentino, Italy
\and Dip. Di Scienze e Innovazione Tecnologica, Univ. del Piemonte Orientale, Viale Teresa Michel, 11, 15121 Alessandria, Italy
\and INAF-Osservatorio Astronomico di Padova, Vicolo dell'Osservatorio 5, 35122 Padova, Italy
\and INAF-Osservatorio Astronomico di Trieste, Via G.B. Tiepolo 11, 34143 Trieste, Italy
\and INAF-Osservatorio Astronomico d'Abruzzo, Via Mentore Maggini, 64100 Teramo, Italy
\and IMCCE, Observatoire de Paris, PSL Research University, CNRS, Sorbonne Universit\'e s, UPMC, France
\and GEOPS, Univ. Paris-Sud, CNRS, Univ. Paris-Saclay, Orsay, France
\and Dipartimento di Ingegneria Civile, Edile ed Ambientale, Univ. La Sapienza, Roma, Italy
\and Italian Meteor and TLE Network (IMTN), Italy
\and IMPMC, Mus\'eum National d'Histoire Naturelle, Paris, France
\and LAM, Institut Pytheas, Marseille, France
\and CEREGE, Institut Pytheas, Marseille, France
\and FRIPON, Collaborative Team, Paris, France
}
\date{Received: date / Revised version: date}
%
\abstract{
On May 30th, 2017 at about 21h 09m 17s UTC a green bright fireball crossed the sky of north-eastern Italy. The fireball path was observed from some all-sky cameras starting from a mean altitude of $81.1 \pm 0.2$ km (Lat. $44.369^{\circ} \pm 0.002^{\circ}$ N; Long. $11.859^{\circ} \pm 0.002^{\circ}$ E) and extinct at $23.3 \pm 0.2$  km (Lat. $45.246^{\circ} \pm 0.002^{\circ}$ N; Long. $12.046^{\circ} \pm 0.002^{\circ}$ E), between the Italian cities of Venice and Padua. In this paper, on the basis of simple physical models, we will compute the atmospheric trajectory, analize the meteoroid atmospheric dynamics, the dark flight phase (with the strewn field) and compute the best heliocentric orbit of the progenitor body. Search for meteorites on the ground has not produced any results so far.
\PACS{
      {PACS-key}{discribing text of that key}   \and
      {PACS-key}{discribing text of that key}
     } 
} 
\authorrunning{Carbognani et al.}
\titlerunning{The May 30th, 2017 Italian Fireball}
\maketitle
\section{Introduction}
\label{sec:intro} 

One of the most interesting astronomical phenomenon that can be seen in the sky is a fireball, namely a very bright meteor caused by the impact of a big meteoroid into the atmosphere.\footnote{According to the IAU definition, a bolide or a fireball is a meteor brighter than absolute visual magnitude -4 (distance of 100 km).} 
Unfortunately, as the events are sporadic and unpredictable, it is not possible to know when you will see the next fireball so we need constant monitoring of the whole sky in order to observe one.\\
A fireball with absolute mag lower than -17 is called superbolide. For small asteroids of tens of meters in diameter, a superbolide can be brighter than the Sun when seen from the Earth. An example of 
such an extreme event is the small asteroid of $19.8 \pm 4.6$ meters in diameter exploded at an altitude of about 27 km above the city of Chelyabinsk (Russia) on February 15th, 2013 at about 03:20.5 
UTC \cite{Popova2013}.\\
Often, less cohesive meteoroids during fall, are fragmented into several blocks each of which becomes an independent fireball. A similar event occurred for the bolide seen from Peekskill (New York State) on the evening of October 9th, 1992 at 23:48 UTC. At a height of about 41.5 km extensive fragmentation of the meteoroid occurred. The meteorite recovered at Peekskill was subsequently identified as an H6 breccia meteorite \cite{Beech1995}.\\
If the meteoroid is large enough, and the speed is not too high, a portion of it can survive the atmospheric ablation phase. When the velocity in the atmosphere drops below about 3 km/s the mass loss and the radiation
emission cease and the meteoroid enters the dark flight phase \cite{Ceplecha1998}. A process of surface cooling begins and the trajectory of the body becomes more and more vertical. The impact velocity 
of the meteoroid on the Earth's surface ranges from 10 - 100 m/s for bodies of mass between 10 g - 10 kg and geocentric speed of about 15 km/s \cite{Ceplecha1998}. What remains of the meteoroid on the
ground is called a meteorite. Most meteoroids totally disintegrate in the air and the impact of some fragments with the Earth's surface is rare.\\
Meteorites are very important because they provide information on the composition and thermal history of asteroids in the early Solar System, and provide a possible vehicle for the dissemination of 
water and organic materials. For these reasons it is important to recover as many meteorites as possible after observation of  bright fireballs events.\\
The physical analysis of a fireball event can be ideally divided into four distinct phases:

\begin{enumerate}
\item Triangulation between different stations on the ground for the reconstruction of the average fireball trajectory in the atmosphere.
\item Estimate of pre-atmospheric velocity, mean drag/ablation coefficients and mass-section ratio. From pre-atmospheric velocity, correcting it for the Earth's attraction and rotation (with 
``zenith attraction'' method), we can compute the true geocentric velocity.
\item Starting from the terminal point of the luminous path, modeling of the dark flight phase to estimate the area on the ground where to look for possible meteorites (strewn field).
\item Compute the heliocentric velocity from the meteoroid true geocentric velocity and, knowing the position vector of the Earth at the fireball time, compute the meteoroid heliocentric osculating orbit. 
The knowledge of the heliocentric orbit is important because it allows to go in search of the meteoroid progenitor body.
\end{enumerate}   

This is the logical path we will follow in this paper, applied to the Italian fireball of May 30, 2017 which we will also call with the code ``country code yyyymmdd'', i.e. IT20170530.

\section{PRISMA, FRIPON, IMTN and CMN networks}
\label{sec:networks} 

\begin{figure}
\begin{center}
\resizebox{0.60\textwidth}{!}{\includegraphics{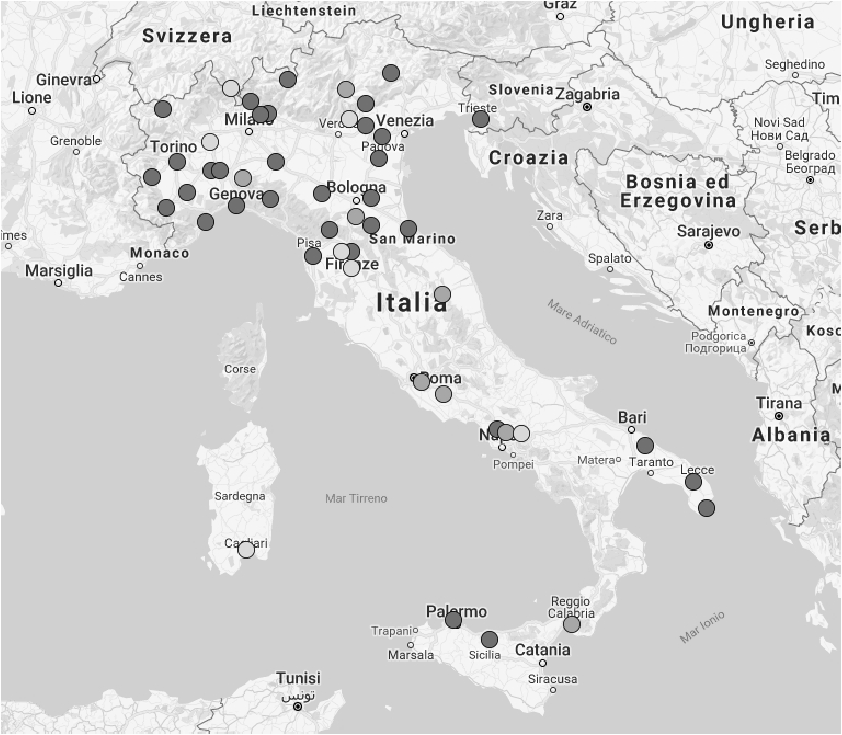}}
\end{center}
\caption{A map from Google Earth showing the distribution of PRISMA stations on April, 2019. Most of the stations are concentrated in northern Italy. 
Legend: dark dots = working camera; grey/white dots = setup phase.}
\label{fig:PRISMA_stations_20190319}
\end{figure}

\begin{figure}
\begin{center}
\resizebox{0.60\textwidth}{!}{\includegraphics{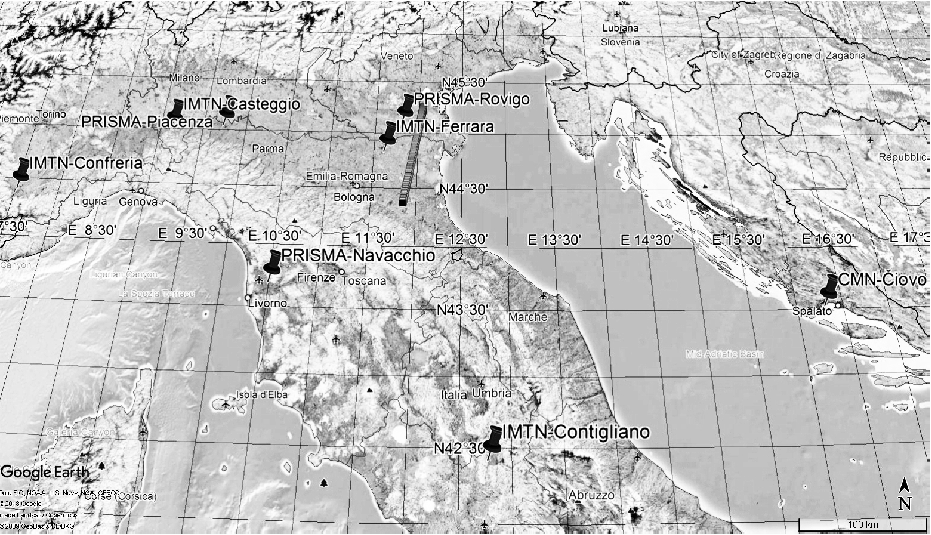}}
\end{center}
\caption{A Google Earth map showing the position of the stations listed in Table~\ref{tab:stations_table} and the fireball trajectory projected on the ground.}
\label{fig:Stations_distribution}
\end{figure}

PRISMA\footnote{www.prisma.inaf.it} network was born in 2016 \cite{Gardiol2016} and means ``Prima Rete Italiana per la Sorveglianza sistematica di Meteore e Atmosfera'', i.e. First Italian Network for Meteor and Atmosphere systematic Surveillance, and is managed by INAF, the italian National Institute for Astrophysics. The word ``first'' in the abbreviation of PRISMA indicates the first national network dedicated to fireballs. In Italy there are also local amatorial networks for observing meteors and in fact, for the pourpose of this paper, we used data from one of them. PRISMA's primary goal is to observe fireballs and recovery any subsequent meteorites while progressively increasing the number of all-sky automatic cameras throughout Italy, so as to have a camera every 80-100 km. As Italy has a surface of 301,338 $\textrm{km}^2$, we need about 50 cameras to cover the whole country with squares of 80 km side. It is crucial to note that inclusion in the PRISMA network is on a voluntary basis. Anyone can cooperate (universities, research centers, schools, amateur astronomers and so on), but must find funding for the purchase of the station's hardware. At present (Nov 2019), 51 all-sky cameras are available, 37 devices in full working mode and 14 in setup phase (see Fig.~\ref{fig:PRISMA_stations_20190319}), whereas on May 2017 only five PRISMA cameras were in operation, mostly in northern Italy.\\
The PRISMA project is an international European collaboration with the French project FRIPON\footnote{www.fripon.org} (Fireball Recovery and InterPlanetary Observation Network), started in 2014 and managed by l'Observatorire de Paris, Mus\'eum National d'Histoire Naturelle, Universit\'e Paris-Sud, Universit\'e Aix Marseille and CNRS \cite{Colas2015}. 
The hardware (and software) of a PRISMA station is similar to a FRIPON station and consists of a small CCD camera kept at room temperature and equipped with a short focal lens objective in order to 
have a wide FoV (Field of View). The camera is connected via LAN to a local mini-PC with Linux-Debian operating system and equipped with a mass storage device of about 1 TB. \\
Acquisition and detection are done on the local mini-PC by open source software FreeTure\footnote{github.com/fripon/freeture}, i.e. Free software to capTure meteors \cite{Audureau2015}, developed by 
the FRIPON team. Acquisition rate is 30 fps and only bright events with a negative magnitude are recorded. For event detection FreeTure uses the subtraction of two consecutive frames with a detection threshold. In order to reduce the amount of false positives, before starting a detection FreeTure waits for a third frame with something moving in the FoV. Every time something bright passes through the FoV of the camera there is a detection and all the images regarding the event, in standard FITS format, are saved in the HDD of the mini-PC that controls the camera. A message is sent to the central FRIPON server, located in the Laboratoire d'Astrophysique de Marseille (LAM), for each local detection. If there is a simultaneous detection in another location the data are downloaded; if not, the data stay on the local HDD for 2 months before being deleted. The reduction pipeline on LAM is launched for every multiple detection.\\
Regarding data reduction, every 10 minutes a calibration image with an exposure time of 5 s is taken from station. In these calibration images stars up to the apparent mag +4.5 are used for the 
astrometric calibration of the camera using the pipeline on LAM developed by FRIPON team and based on the software SExtractor and SCAMP\footnote{www.astromatic.net/software}. SExtractor is a program 
that builds a catalogue of objects from an astronomical image, while SCAMP reads SExtractor catalogs and, using a star catalog, computes astrometric and photometric solutions for any arbitrary sequence 
of FITS images in a completely automatic way \cite{Bertin1996}, \cite{Bertin2006}. As an astrometric catalog SCAMP uses the Tycho 2 star catalog. For the known stars the accuracy of astrometric
calibration is between 100 and 200 arcsec root mean square. More frequently around 150 arcsec.\\
The IMTN\footnote{meteore.forumattivo.com} (Italian Meteor and TLE Network) is a surveillance network managed by amateur astronomer both for the study of meteors and high-atmosphere phenomena or TLE, Transient Luminous Events. Generally IMTN stations have a smaller FoV than PRISMA stations, because the camera objective tends to have higher focal lengths but, on the other hand, 
the images have slightly higher resolution. An observation from a station of the Croatian Meteor Network (CMN) was also collected. The CMN\footnote{cmn.rgn.hr} consists of 30 surveilance cameras 
each having a FoV of $64 \times 48$ deg. The cameras monitor most of the night sky over Croatia. See Table~\ref{tab:stations_table} and Fig.~\ref{fig:Stations_distribution} for more details about the cameras that captured IT20170530. IMTN members normally use commercial software as UFOOrbit given by SonotaCo\footnote{sonotaco.com} for the movie capture, the astrometric reduction of the fireballs 
images and the computation of the trajectory and orbits. As astrometric star catalog IMTN uses the SKYMAP Master Catalog\footnote{tdc-www.harvard.edu/catalogs/sky2k.html}, Version 4, which features 
an extensive compilation of information on almost 300,000 stars brighter than 8.0 mag. In this case the accuracy of astrometric calibration is between 100 and 180 arcsec root mean square.

\begin{table*}
\centering
\caption{Some technical data about PRISMA, IMTN and CMN stations that observed IT20170530. From left to right: name, latitude, longitude and elevation over sea level; camera model, objective focal 
length, approximate field of vision, frame rate and scale of images in arcsec/pixel. In bold type the stations whose observations were used to compute the fireball trajectory (see text about this choice).}
\label{tab:stations_table}

\begin{tabular}{lcccccccc}
\hline
Name & Lat. N ($^\circ$) & Long. E ($^\circ$)& El. (m) & Camera model & FL (mm) & FoV ($^\circ$) & fps & Scale\\
\hline
PRISMA-Navacchio          &	43.68320 & 10.49163 & 015 & Basler A1300gm       & 1.2 & $223 \times 166$ & 30 & 600\\
PRISMA-Piacenza           & 45.03538 & 09.72503 & 077 & Basler A1300gm       & 1.2 & $223 \times 166$ & 30 & 600\\
\textbf{PRISMA-Rovigo}             & 45.08167 & 11.79505 & 015 & Basler A1300gm       & 1.2 & $223 \times 166$ & 30 & 600\\
\textbf{IMTN-Casteggio}            & 44.98810 & 09.12510 & 238 & Mintron MTV-12V6H-EX & 4.0 & $92 \times 69$ & 25 & 430\\
IMTN-Confreria            & 44.39590 & 07.51730 & 550 & Watec 120 N+         & 4.5 & $82 \times 62$   & 25 & 385\\
\textbf{IMTN-Contigliano}	      & 42.41140 & 12.76820 & 421 & Mintron MTV-12V6H-EX & 4.0 & $92 \times 69$ & 25 & 430\\
IMTN-Ferrara              & 44.81760 & 11.61060 & 009 & Mintron MTV-12V6H-EX & 2.6 & $142 \times 106$ & 25 & 666\\
CMN-Ciovo (Croatia)       & 43.51351 & 16.29545 & 020 & SK1004X              & 4.0 & $64 \times 48$ & 25 & 325\\
\hline
\end{tabular}
\end{table*}

\section{Software packages for fireballs analysis}
\label{sec:software} 
We did not use the FRIPON astrometric pipeline that we briefly illustrated above for the astrometric reduction of the fireball images. Rather, for the PRISMA team the observation of IT20170530 was a 
good opportunity to start developing an independent pipeline. In this section we briefly describe the main software packages available so far.

\subsection{Astrometry}
The determination of an astrometric solution for all-sky cameras has been already discussed in literature, from Ceplecha to Borovi\v cka \cite{Ceplecha1987, Borovicka1992, Borovicka1995}. The astrometric model is based on a parametric description of heavy optical distortions in the radial coordinate due to the lens type \cite{Ceplecha1987}. Minor but still significant effects due to the displacement of the optical axis with respect to the zenith direction and camera misalignments are taken into account in a refined model \cite{Borovicka1992, Borovicka1995}. We have implemented this model with IDL (Interactive Data Language)\footnote{www.harrisgeospatial.com/SoftwareTechnology/IDL.aspx} to derive the astrometry of the fireball by means of the IDL-Astro and Markwardt-IDL libraries. On a bright fireball the position error on a single frame is of the order of 1 arcmin, so that any astrometric inaccuracies introduced by the model become negligible using a set of calibration data spanning a few months of observations. This is not completely true at very low elevation, especially with a degraded point spread function (PSF). Regarding the details about the astrometric reduction technique see Barghini et al., 2019 \cite{Barghini2019}.

\subsection{Physical analysis}
Analysis of the astrometry data files from IDL was carried out by writing a software able to run under MATLAB\textsuperscript{\textregistered} Release 2015b (MathWorks\textsuperscript{\textregistered}, Inc., Natick, Massachusetts, United States)\footnote{http://www.mathworks.com}. The code has been divided into four main functions concerning trajectory, dynamics model, dark flight (with strewn field) and osculating heliocentric (or barycentric) orbit. In the trajectory function the preliminary atmospheric path of the fireball was computed as geometric intersection of the best planes containing two stations and the unit vectors of the fireball's observed points \cite{Ceplecha1987}. The definitive best fireball trajectory was obtained with observations from $N > 2$ stations simultaneously, using the Borovi\v cka method \cite{Borovicka1990}. In this way the values of the starting and terminal height ($H_s$ and $H_t$), of the trajectory inclination on the Earth's surface ($T_i$), and of the azimuth were the fireball came from ($A_z$) are found. Associated with the heights values there are also the geographical coordinates, longitude and latitude, respectively starting ($Lo_s$, $La_s$) and terminal 
($Lo_t$, $La_t$). Following this method the result about fireball trajectory is always a straight line and it is interesting to note that it is not necessary to have accurate temporal data from all the stations to have geometric triangulation. The analysis software package was called MuFiS (Multipurpose Fireball Software) because it includes triangulation, dynamics, dark flight and orbit functions. 
We will see the underground physics of the other MuFiS's functions in the next sections, coupled with the IT20170530 analysis. \\
MuFiS has been verified by applying it to a synthetic fireball with known initial parameters. The synthetic fireball was generated by writing a completely independent software. To be realistic, four trajectories seen from four different stations have been simulated and, on each trajectory, a random uncertainty of 1/100 s over time and 1 arcmin in the position was added. In general, the agreement between the synthetic data and those found by MuFiS is very good, see Table~\ref{tab:trajectory_synthetic} for a comparison between synthetic and MuFiS values. It is interesting to note that, if you adopt a drag coefficient value ($\Gamma$) different from the one used to generate the synthetic fireball, the triangulation values are the same but the guess estimated values about the meteoroid mass and diameter change. This happens because the dynamic model fit provides the $\Gamma / D_\infty$ ratio only, where $D_\infty$ is the meteoroid pre-atmospheric ratio mass/section. Fortunately, the terminal point values from dynamical analysis (height $H_{fin}$, velocity $v_{fin}$ and acceleration $a_{fin}$), useful to compute the dark flight phase, are independent from the $\Gamma$ value as we will see better in Section~\ref{sec:dynamic} and Section~\ref{sec:dark}. The adopted dynamic model will be discuss in detail in Section~\ref{sec:dynamic}. MuFiS detailed structure will be explained in a next future paper.

\section{The fireball atmospheric trajectory}
\label{sec:trajectory} 

\begin{figure}
\begin{center}
\resizebox{0.60\textwidth}{!}{\includegraphics{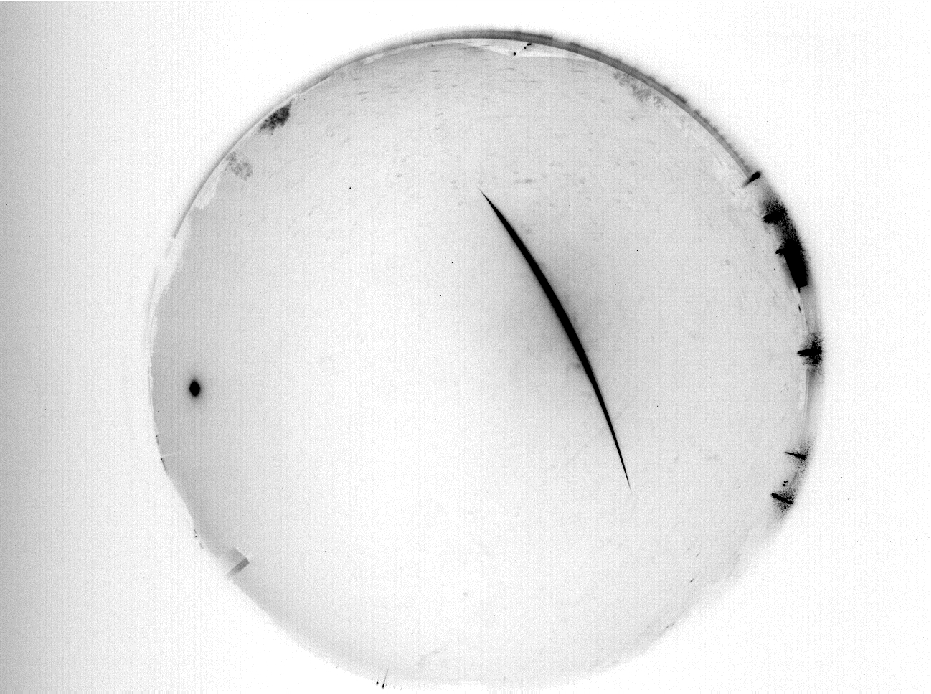}}
\end{center}
\caption{A negative image showing the full path of IT20170530 from Rovigo. North is down, south is up. The bright object on the left is the Moon near the western horizons. The 
fireball moved from top-left to bottom-right. The total duration of the fireball was about 9.51 s. From this image no significant fragmentation of the meteoroid appears.}
\label{fig:Fireball_IT20170530_Rovigo_ITVE02}
\end{figure}

\begin{figure}
\begin{center}
\resizebox{0.60\textwidth}{!}{\includegraphics{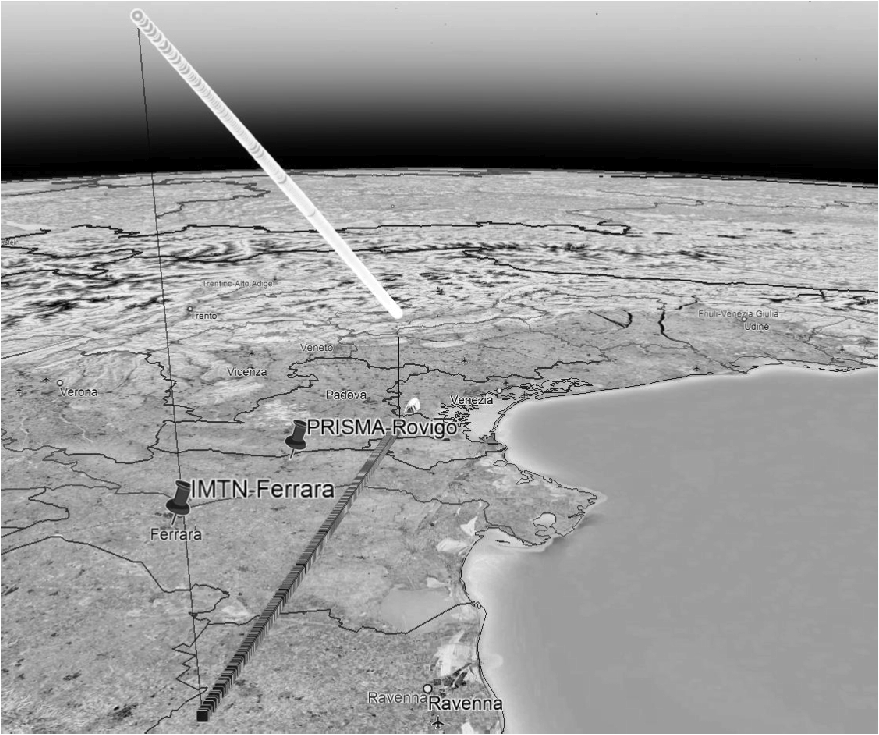}}
\end{center}
\caption{A Google Earth map showing the best trajectories, both in atmosphere and projected on the ground, resulting from the triangulation between PRISMA-Rovigo, IMTN-Casteggio and IMTN-Contigliano.}
\label{fig:IT20170530_trajectories}
\end{figure}

And now let's start with the analysis of the fireball trajectory in the atmosphere. There are 285 position points available from Rovigo, for a total duration of about 9.51 seconds. Unfortunately for Piacenza and Navacchio the bolide was very low above the horizon in an area of the focal plane where resolution is very poor. This, combined with the remarkable distance between both cameras and the bolide (roughly 200 km) makes the determination of the bolide position in geocentric coordinates much more difficult. Fortunately the data from Rovigo are the best of the trio because the fireball passed near the zenith and, thanks to the favorable geographical position, the trajectory terminal point was imaged (see Fig.~\ref{fig:Fireball_IT20170530_Rovigo_ITVE02}). For these reasons, the triangulation of the fireball trajectory was performed with the data from Rovigo crossed with the data from the IMTN/CMN stations. Data from these stations are good for geometrical triangulation (i.e. good astrometry), but hardly reliable for the measurement of the fireball speed. This is because the temporal data of the IMTN/CMN frames is not directly accessible from the commercial software used by IMTN/CMN operators. For this reason, the speed of the fireball were obtained from Rovigo data only. The timing data of a PRISMA station are easily accessible and synchronization is made via NTP protocol, accuracy is not directly measurable but it is better than 100 ms.\\
From the triangulations with the Rovigo station we have excluded the Ferrara station because the angle value between the planes identified by these two stations is only $2.5^\circ$, too 
small to produce a reliable triangulation. The four possible 2-stations atmospheric trajectories are therefore given by Rovigo plane intersected with Casteggio/Confreria/Contigliano and 
\v Ciovo planes. The Rovigo-Casteggio and Rovigo-Contigliano trajectories are practically coincident (the difference is about 0.2 km on the ground), while between the other two remaining 
trajectories there is a difference of about 2 km if projected on the ground. The mean value of the two first trajectories provides the same parameters, but with less uncertainty, than the average 
of all the four trajectories. This makes the trajectory determined using the Borovi\v cka method between Rovigo, Casteggio and Contigliano the best trajectory, with the uncertainty computed as the mean observed deviation from the mean path (see Fig.~\ref{fig:IT20170530_trajectories} and Table~\ref{tab:trajectory_table}).

\begin{table*}
\centering
\caption{This table shows the comparison between a synthetic fireball generated by a 3.5 kg meteoroid (supposed to be a chondrite with an average density of 3500 $\textrm{kg}/\textrm{m}^3$) and ablation coefficient 0.0070 $\textrm{s}^2/\textrm{km}^2$ that falls into the atmosphere with a pre-atmospheric speed of 21 km/s and the solutions found by MuFiS about triangulation and dynamic model. Note that if you adopt a drag coefficient value ($\Gamma$), different from the one used to generate the synthetic fireball, the triangulation values are the same but the guess estimated values about the meteoroid mass and diameter change. The terminal point values, necessary to compute the dark flight phase, is independent from the $\Gamma$ value. For details and limits about the adopted dynamic model see Section~\ref{sec:dynamic}.}
\label{tab:trajectory_synthetic}

\begin{tabular}{lcccc}
\hline
Quantity                                    & Synthetic ($\Gamma = 0.7$)           & MuFiS ($\Gamma=0.70$) & MuFiS ($\Gamma=0.58$) & MuFiS ($\Gamma=0.80$)\\
\hline
$H_s$ (km)                                  & 71.0                                 &  70.8                 &  70.8                 &  70.8\\
$La_s$ (N, $^\circ$)	                    & 44.3694                              &  44.3695              &  44.3695              &  44.3695\\
$Lo_s$ (E, $^\circ$)	                    & 11.8594                              &  11.8600              &  11.8600              &  11.8600\\
\hline
$H_t$ (km)                                  & 21.8                                 &  21.9                 &  21.9                 &  21.9\\
$La_t$ (N, $^\circ$)	                    & 44.7310                              &  44.7310              &  44.7310              &  44.7310\\
$Lo_t$ (E, $^\circ$)	                    & 11.9489                              &  11.9492              &  11.9492              &  11.9492\\
\hline
$T_i$	($^\circ$)                          & 50                                   &  49.8                 &  49.8                 &  49.8\\
$A_z$   (from North to South, $^\circ$)     & 190                                  &  190                  &  190                  &  190\\
\hline
$v_\infty$ (km/s)                           & 21.0                                 &  20.9                 &  20.9                 &  20.9\\
$\sigma$ ($\textrm{s}^2/\textrm{km}^2$)     & 0.0070                               &  0.0068               &  0.0068               &  0.0068\\
$D_\infty$ ($\textrm{kg}/\textrm{m}^2$)	    & 289                                  &  281                  &  233                  &  321\\
$d_\infty (\textrm{m})$                     & 0.12                                 &  0.12                 &  0.10                 &  0.14\\		
$m_\infty (\textrm{kg})$	                & 3.5                                  &  3.2                  &  1.8                  &  4.8\\		
$D_{fin} (\textrm{kg}/\textrm{m}^2$)        & 173                                  &  172                  &  143                  &  197\\ 		
$d_{fin} (\textrm{m})$                      & 0.07	                               &  0.07                 &  0.06                 &  0.08\\	
$m_{fin} (\textrm{kg})$                     & 0.7	                               &  0.7                  &  0.4                  &  1.10\\	
\hline			
$v_{fin}$ (km/s)	                        & 3.0                                  &  3.0                  &  3.0                  &  3.0\\		
$a_{fin}$ ($\textrm{km}/\textrm{s}^2$)      & -2.4 	                               &  -2.4                 &  -2.4                 &  -2.4\\	
$H_{fin}$ (km)	                            & 21.8                                 &  21.9                 &  21.9                 &  21.9\\ 	
\end{tabular}
\end{table*}

\begin{table}
\centering
\caption{Data about the starting/terminal points and the radiant (geocentric apparent and true), of the fireball trajectory from triangulations between PRISMA-Rovigo and IMTN-Casteggio/Contigliano stations.}
\label{tab:trajectory_table}

\begin{tabular}{lc}
\hline
Quantity & Numerical value\\
\hline
$H_s$	                                            & $81.1 \pm 0.2$ km\\
$La_s$ (N, $^\circ$)	                            & $44.369 \pm 0.002$ ($\pm 0.2$ km)\\
$Lo_s$ (E, $^\circ$)	                            & $11.859 \pm 0.002$ ($\pm 0.2$ km)\\
\hline
$H_t$	                                            & $23.3 \pm 0.2$ km\\
$La_t$ (N, $^\circ$)	                            & $45.246 \pm 0.002$ ($\pm 0.2$ km)\\
$Lo_t$ (E, $^\circ$)	                            & $12.046 \pm 0.002$ ($\pm 0.2$ km)\\
\hline
$T_i$	($^\circ$)	                                & $30.8^\circ \pm 0.1^\circ$\\
$A_z$   (from North to South, $^\circ$)             & $188.7^\circ \pm 0.1^\circ$\\
\hline
$\alpha_{GAR}^{a}$ (J2000.0)                            & $209.9^\circ \pm 0.1^\circ$\\
$\delta_{GAR}^{a}$ (J2000.0)                            & $-14.5^\circ \pm 0.1^\circ$\\
\hline
$\alpha_{GTR}^{b}$ (J2000.0)	                        & $207.4^\circ \pm 0.2^\circ$\\
$\delta_{GTR}^{b}$ (J2000.0)	                        & $-25.4^\circ \pm 0.6^\circ$\\
\hline
\\
\multicolumn{2}{l}{$^{a}$ GAR = Geocentric Apparent Radiant.}\\ 
\multicolumn{2}{l}{$^{b}$ GTR = Geocentric True Radiant.}\\ 
\end{tabular}
\end{table}

The observed fireball path begins from a starting height $H_s = 81.1 \pm 0.2$ km and extinct at a terminal height $H_t = 23.3 \pm 0.2$ km, between the Italian cities of Venice and Padua. The total length of the luminous atmospheric path is about 115 km (Fig.~\ref{fig:Height_vs_time}). With these height values above sea level the observed path was in a continuum flow regime and this affected the choice of the physical model to describe the fall of the meteoroid into the atmosphere \cite{Campbell-Brown2004}.\\ 
The intersection between the fireball trajectory with the celestial sphere gives the apparent radiant position. Correcting it for the Earth's rotation and gravity (using ``zenith attraction'' method), 
we can compute the position of the true radiant \cite{Ceplecha1987}. The apparent radiant is in Virgo constellation, the true radiant in Hydra. The resulting values from triangulation are shown in 
Table~\ref{tab:trajectory_table}. The rectangular geocentric coordinates $x_i$, $y_i$ and $z_i$ for every observed trajectory points are given directly by triangulation while the mean velocity between 
the observed points can be, as a first approximation, computed using the Pythagorean theorem:

\begin{equation}
    v_i\approx\frac{\sqrt{(x_{i+1}-x_i)^2+(y_{i+1}-y_i)^2+(z_{i+1}-z_i)^2}}{(t_{i+1}-t_i)}.
	\label{eq:pytgagorean}
\end{equation}

However this simple numerical approximation, given by Eq.~(\ref{eq:pytgagorean}), for the first derivative of the positions lead to more uncertainty which we reduced computing the central difference
between data points. The central difference approximation is more accurate for smooth functions, as in our case. Things get worse if we compute the acceleration in the same kinematic way. 
However, it is important to know the speed instant by instant to trace the pre-atmospheric speed, before the entry of the meteoroid into the atmosphere, and the terminal speed that precedes the (possible), dark flight phase. So in order to compute the best height, velocity and acceleration (or mass-section ratio) in the terminal point of the luminous path, crucial parameters to the dark flight phase model, we have used a meteoroid single body dynamic model in a continuous flow regime to fit the observed data of heights and speeds vs time. This model will be extensively discuss in the next section.

\section{The physical dynamic model of the meteoroid}
\label{sec:dynamic} 

In order to estimate the fireball main physical parameters, i.e. drag and ablation coefficients, pre-atmospheric velocity, mass/section ratio and to compute the best height, velocity and acceleration 
in the terminal point of the luminous path, we have used a single body dynamical model numerically integrating the differential equations describing the motion and the ablation of the meteoroid. 
In this classical model no fragmentation is taken into account, but for our fireball no significant fragmentation of the meteoroid appears, see 
Fig.~\ref{fig:Fireball_IT20170530_Rovigo_ITVE02}. Ablation begins when the surface of the meteoroid reaches the boiling temperature. At this point the temperature is assumed to remain constant and the light emission negligible respect to the kinetic energy of the meteoroid \cite{Campbell-Brown2004}. Under the hypothesis of a constant ablation rate and a constant body shape during its ablation, ours starting differential equations are as follows \cite{Kalenichenko2006}:

\begin{equation}
    \frac{dv}{dt} = -\frac{\Gamma\rho_a v^2}{D_\infty}\exp\left({-\frac{\sigma}{6}\left(v^2-{v_\infty}^2\right)}\right)
	\label{eq:velocity}
\end{equation}

\begin{equation}
    \frac{d\rho_a}{dt} = \frac{\rho_a v \cos z}{H}
	\label{eq:density}
\end{equation}

\begin{equation}
    \frac{dh}{dt}= -v \cos z
	\label{eq:height}
\end{equation}

Eq.~(\ref{eq:velocity}) comes from the momentum-energy conservation, Eq.~(\ref{eq:density}) is a consequence of the simple atmospheric density model adopted (i.e. the 1976 U. S. 
standard atmosphere model fitted with an exponential function), and Eq.~(\ref{eq:height}) expresses a straight-line model, i.e. we neglect the meteor curvature, which is a reasonable assumption for short trajectories of the order of about 100 km in length as in our case \cite{jeanne2019}. In the previous equations the symbols have the meaning listed in Table~\ref{tab:symbol1_table}.

\begin{table}
\centering
\caption{Meaning of the symbols for Eq.~(\ref{eq:velocity}), (\ref{eq:density}) and (\ref{eq:height}).}
\label{tab:symbol1_table}
\begin{tabular}{ll}
\hline
Symbol & Quantity \\
\hline
$v$ & Body speed with respect to the air\\
$\Gamma$ & Aerodynamic drag coefficient\\
$\rho_a$ & Air density (from 1976 U. S. standard atmosphere model)\\
$\sigma$ & Ablation coefficient\\
$v_\infty$ & Pre-atmospheric velocity\\
$D_\infty = m_\infty/A_\infty$ & Pre-atmospheric mass/cross section ratio\\
$z$ & Mean zenith distance of the fireball radiant\\
$H$ & Effective atmosphere scale height (from 1976 U. S. standard atmosphere model)\\
\hline
\end{tabular}
\end{table}

Note that the aerodynamic drag coefficient $\Gamma$ is equal to $C_d/2$, where $C_d$ is the usual drag coefficient used in fluid dynamics \cite{Kalenichenko2006}. 
Numerically integrating the previous differential equations with Runge-Kutta 4th/5th order methods and comparing the result with the observed values of $h(t)$ and $v(t)$ it was possible to estimate 
the best value of  $\sigma$, $v_\infty$ and $D_\infty$ that fit the observed value starting from appropriate guess values. So we had performed a multi-parameter fitting using the observed data from trajectory and velocity together. The drag coefficient $\Gamma$ has been kept fixed because, from Eq.~(\ref{eq:velocity}), it is coupled to parameter $D_\infty$ and cannot be determined 
separately. This choice weakly affects the subsequent dark flight phase because what matters is the $\Gamma/D_\infty$ ratio (see Table~\ref{tab:trajectory_synthetic}). \\
As a starting value for the pre-atmospheric velocity we have fitted the fireball observed velocity $v$ vs. the height $h$ with the following exponential model from Ceplecha \cite{Ceplecha1961}:

\begin{equation}
    v = v_\infty + c_v e^{\left({-bh}\right)}
	\label{eq:ceplecha_velocity}
\end{equation}

In this equation $c_v$ and $b$ are constants to be determined together with $v_\infty$. In general these phenomenological models tend to be less accurate than the multi-parametric fit but we only needed 
an estimate of the initial pre-atmospheric velocity \cite{Egal2017}. Instead, as a starting value for $D_\infty$, we have taken Eq.~(\ref{eq:velocity}) computed with the estimated quantity in 
the terminal point of the luminous path: 

\begin{equation}
    D_\infty \approx \left(-\frac{\Gamma\rho_a v^2}{dv/dt}\right)_{fin}\exp\left({-\frac{\sigma}{6}\left({v_\infty}^2-{v_{fin}}^2\right)}\right)
	\label{eq:velocity_guess}
\end{equation}

\begin{figure}
\begin{center}
\resizebox{0.60\textwidth}{!}{\includegraphics{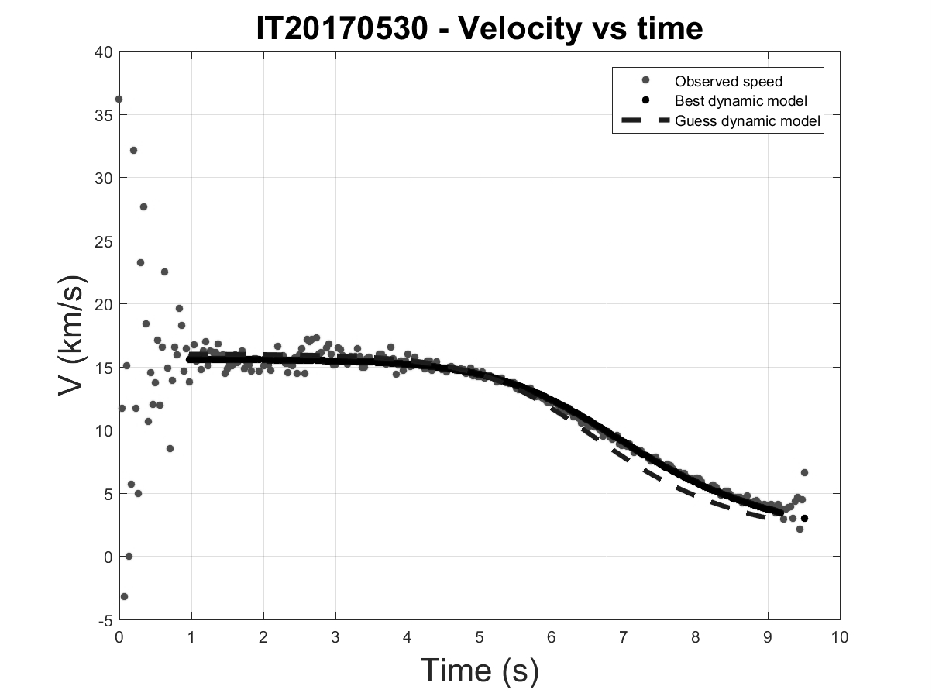}}
\end{center}
\caption{The fireball velocity vs. time as a result of the triangulation from Rovigo, Contigliano and Casteggio. 
The speed is computed using Rovigo's temporal data only. The initial dispersion of the points is due to the fact that the fireball was very far from the station and the displacement was low. In this condition the sky position uncertainty is the dominant factor in the computed velocity. Grey dots = observed values; dotted black line = model with starting guess values; black line = best fit model.
}
\label{fig: Velocity_vs_time}
\end{figure}

In Eq.~(\ref{eq:velocity_guess}) the guess values of the fireball final velocity and the final acceleration are also estimated from the Ceplecha kinematic model (Eq.~(\ref{eq:ceplecha_velocity})). 
We also have put $\sigma \approx 0.006$ $\textrm{s}^2/\textrm{km}^2$, the mean of the typical intrinsic ablation coefficient \cite{Ceplecha2005} and $\Gamma_{fin} = 0.58$ 
\cite{Ceplecha1987}. This last value for drag coefficient is equal to the starting value because we fixed $\Gamma$. In Section~\ref{sec:dark} we will see the effect of $\Gamma$ variation on the dark flight phase.\\ 
In addition to the parameters that characterize the meteoroid, the solution of the differential equations of motion also depends by the starting height, speed and air density. Height and density values 
are well determined by observations and atmospheric model respectively. The initial speed value $V_{start}$ is more uncertain, see Fig.~\ref{fig: Velocity_vs_time}. As initial guess value of the meteoroid velocity we took the one given by the Ceplecha model fit of the observed data (Eq.~(\ref{eq:ceplecha_velocity})). These initial guess values were allowed to vary in an appropriate 
physical range (see column 4 of Table~\ref{tab:dynamic_table}).\\
After a least square fit of the $h(t)$ and $v(t)$ observed values, we obtained the results given in Table~\ref{tab:dynamic_table}. 
The integration time start 1 s after the fireball first detection from Rovigo and ends at 9.2 s, about 0.3 s before the end of the fireball path. In this way we exclude the noisiest observed points 
from the numerical integration (see Fig.~\ref{fig: Velocity_vs_time}). 

\begin{table*}
\centering
\caption{Values about the best fit parameter of the dynamic model, guess values and allowed variation range ($\Gamma$ is keep fixed with a guess value equal to 0.58, as in 
\cite{Ceplecha1987}. In the last three rows of the table are the values of height, velocity and acceleration in the terminal point (t = 9.51 s), given by the model. Thanks to the low atmospheric speed the meteoroid ablation was only partial and the model tells us that it is possible to find a small meteorite. Pay attention that the fireball physical model provides the $\Gamma/D_\infty$ ratio only. The pre and post-atmospheric mass and diameter values are guess assuming a mean density of about 3500 kg/$\textrm{m}^3$, i.e. a typical chondrite value, and a spherical shape. Their values are only indicative, not necessarily corresponding to the truth. Note that the $\sigma$ value was blocked from the inferior boundary condition. If the range of variation is widened the ablation value would become even smaller. It is not a software problem because the ablation values of the synthetic fireballs are correctly reproduced (see Table~\ref{tab:trajectory_synthetic}). Perhaps this low ablation coefficient was simply due to a compact rock material.}
\label{tab:dynamic_table}

\begin{tabular}{lccc}
\hline
Quantity & Best Value & Guess Value & Allowed Range (min; max)\\
\hline
$v_\infty$ (km/s) & $15.9 \pm 0.3$ & $16.1$	& $v \pm 0.5$\\
$\sigma$ ($\textrm{s}^2/\textrm{km}^2$) &	$0.0012 \pm 0.0002$ &	0.006 &	$\sigma/5$; $5\sigma$\\
$V_{start}$ (km/s) & $15.6 \pm 0.3$ & 15.9 &	$V \pm 1$\\
$D_\infty$ ($\textrm{kg}/\textrm{m}^2$)	& $234 \pm 15$ & 199 & $D/5$; $5D$\\
$d_\infty^{a}$ ($\textrm{m}$)           & 0.1   &  &   \\		
$m_\infty^{b}$ ($\textrm{kg}$)	        & 1.8   &  &  \\		
$D_{fin}$ ($\textrm{kg}/\textrm{m}^2$)  &	$220 \pm 10$            &       &   \\ 		
$d_{fin}^{c}$ ($\textrm{m}$)            & 0.09	&  &  \\	
$m_{fin}^{d}$ ($\textrm{kg}$            & 1.6	&  &  \\	
\hline			
$v_{fin}$ 	       & $3.0 \pm 0.3$ km/s &  &  \\		
$a_{fin}$          & -1.19 $\textrm{km}/\textrm{s}^2$	    &  &  \\	
$H_{fin}$ 	       & $23.4 \pm 0.1$ km  &  &  \\ 		
\hline
\\
\multicolumn{4}{l}{$^{a}$ pre-atmospheric diameter if chondrite.}\\ 
\multicolumn{4}{l}{$^{b}$ pre-atmospheric mass if chondrite.}\\ 
\multicolumn{4}{l}{$^{c}$ post-atmospheric diameter if chondrite.}\\ 
\multicolumn{4}{l}{$^{d}$ post-atmospheric mass if chondrite.}\\
\end{tabular}
\end{table*}

\begin{figure}
\begin{center}
\resizebox{0.60\textwidth}{!}{\includegraphics{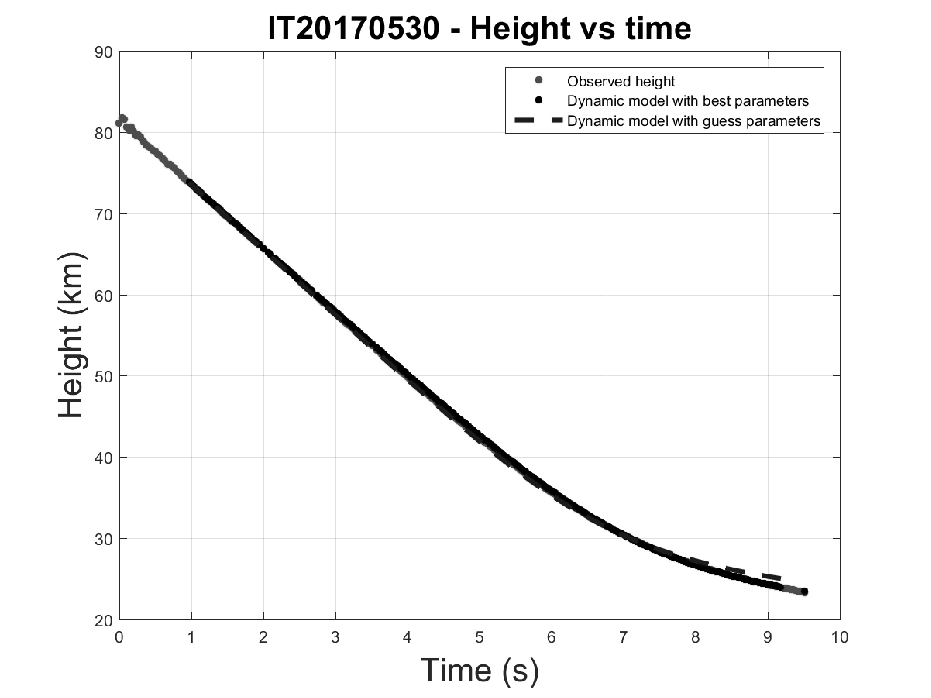}}
\end{center}
\caption{The fireball height vs. time as a result of the triangulation from Rovigo, Contigliano and Casteggio.  
Grey dots = observed values; dotted line = model with starting guess values; black line = best fit model.
}
\label{fig:Height_vs_time}
\end{figure}

\begin{figure}
\begin{center}
\resizebox{0.60\textwidth}{!}{\includegraphics{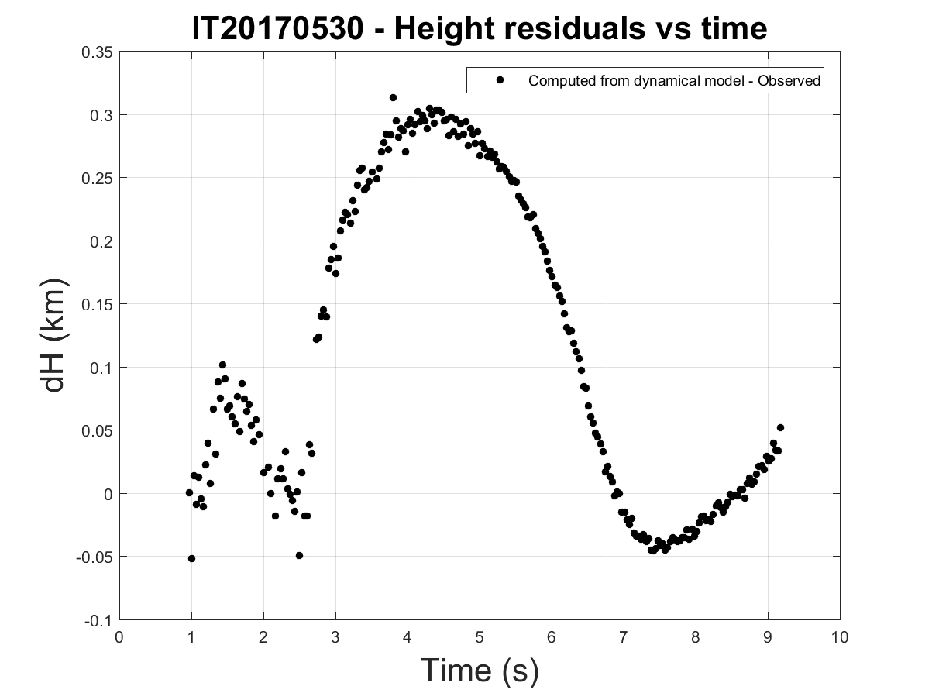}}
\end{center}
\caption{The residual between the observed height vs. time values and the dynamical model of the meteoroid (Rovigo, Contigliano and Casteggio). The mean residual value is about 0.12 km, discussion in the text.}
\label{fig:Height_residuals_vs_time}
\end{figure}

\begin{figure}
\begin{center}
\resizebox{0.60\textwidth}{!}{\includegraphics{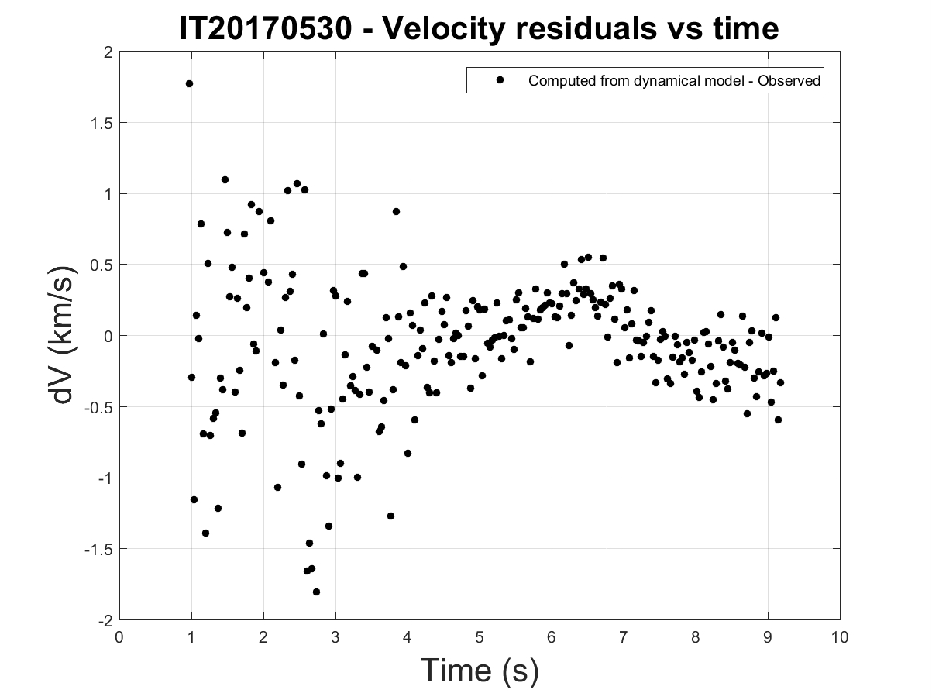}}
\end{center}
\caption{The residual between the observed velocity values vs. time and the dynamical model of the meteoroid (Rovigo with Contigliano and Casteggio). The mean residual value is about 0.3 km/s, discussion in the text.}
\label{fig:Velocity_residuals_vs_time}
\end{figure}

\begin{figure}
\begin{center}
\resizebox{0.60\textwidth}{!}{\includegraphics{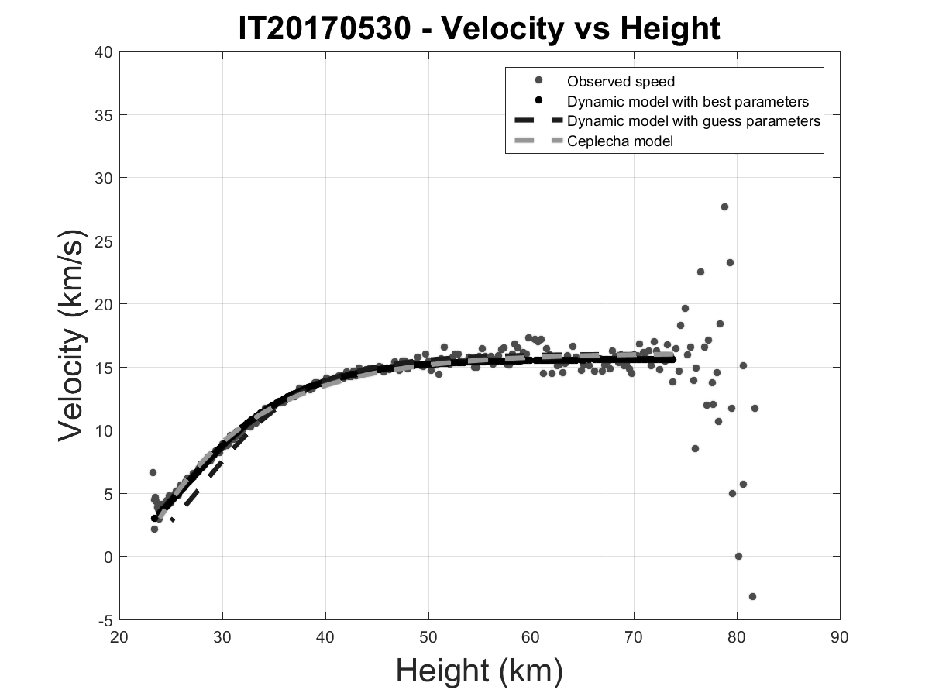}}
\end{center}
\caption{The fireball velocity vs. height as a result of the triangulation from Rovigo with Contigliano and Casteggio.  
Grey dots = observed values; dotted black line = model with starting guess values; black line = best fit model.
}
\label{fig:Velocity_vs_Height}
\end{figure}

\begin{figure}
\begin{center}
\resizebox{0.60\textwidth}{!}{\includegraphics{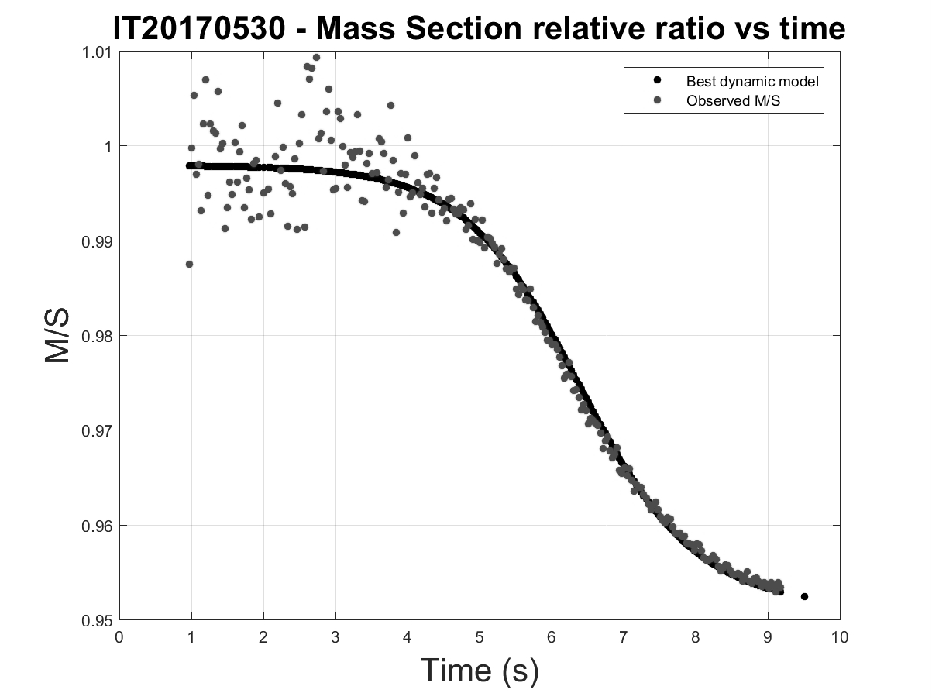}}
\end{center}
\caption{The fireball relative mass/(cross section) ratio vs. time as a result of the triangulation from Rovigo with Contigliano and Casteggio. 
In this figure it is possible to follow the meteoroid ablation vs time because $M/S \propto r$, where $r$ is the meteoroid radius. 
The best fit model line start below 1 because the first observation was made when the ablation was already on. The scattering is due to the initial uncertainty on the fireball speed.
Grey dots = observed values; black line = best fit model.
}
\label{fig:Mass_Section_relative_ratio_vs_time}
\end{figure}

The initial guess values describe the first part of the trajectory very well, between 1 - 5 s, but not the last one (see Fig.~\ref{fig:Height_vs_time}, Fig.~\ref{fig: Velocity_vs_time} and 
Fig.~\ref{fig:Velocity_vs_Height}). So it is the second half of the trajectory that determines the best fit of the free parameters. The mean residuals are about 0.3 km/s for velocity and 0.1 km 
for height (see Fig.~\ref{fig:Height_residuals_vs_time} and Fig.~\ref{fig:Velocity_residuals_vs_time}). In the height residuals there is an evident systematic trend which remains confined within 0.3 km and is below 0.05 km near the terminal trajectory point: there are some height variations that the model cannot completely reproduce. The velocity trend appear better described although we can see a systematic effect between 5 and 8 seconds with an amplitude of about 0.3-0.4 km/s, about the order of magnitude of speed uncertainty. Assuming a mean density of about 3500 kg/$\textrm{m}^3$ we can estimate the mass and dimension of the meteoroid (see Table~\ref{tab:dynamic_table}). With the dynamic model results we can also compute a synthetic fireball lightcurve. Assuming that a fraction 0.04 of the meteoroid kinetic energy is converted into visible radiation we found that the absolute magnitude reached a minimum of about -7.3 about 6.5 s after observation start \cite{Gritsevich2011}. This is a synthetic estimate of absolute magnitude at maximum brightness, so its value must be taken with caution. The computation of the absolute magnitude from Rovigo images is difficult because they are saturated and the values given by IMTN cameras are not reliable for the lack of the temporal data. The pre-atmospheric velocity is $v_\infty = 15.9 \pm 0.3$ km/s. Velocities of solar-system meteoroids at their encounter with the Earth's atmosphere are within the following limits \cite{Ceplecha1998}: the lower one 11.2 km/s, if the meteoroid approach the Earth from behind with zero relative velocity, the upper one 72.8 km/s, if meteoroid struck the Earth head-on. In this last case we add the 42.5 km/s parabolic velocity at Earth's perihelion plus 30.3 km/s, the velocity of the Earth at perihelion. So the meteoroid belonged to the Solar System. Correcting $v_\infty$ for the attraction and the rotation of the Earth we finally obtain the meteoroid geocentric velocity before entering the Earth's atmosphere \cite{Ceplecha1987}: $v_g = 11.4 \pm 0.4$ km/s. 
The corresponding heliocentric velocity is $v_g = 37 \pm 1$ km/s. In Section~\ref{sec:orbit}, knowing the position vector of the Earth at the fireball time, we will compute the meteoroid heliocentric osculating orbit.

\section{The dark flight phase and the strewn field}
\label{sec:dark}
 
In order to model the dark flight phase, it is important to know the profile of the atmosphere in the time and place closest to the meteoroid fall because the residual meteoroid trajectory, after the end of the luminous path, can be heavily influenced by the atmospheric conditions.\\
The data about wind velocity, wind direction, density, pressure and temperature vs. the height above Earth's surface can be obtained from weather balloons up to an altitude of about 30 - 40 km. In 
Italy there are 8 weather stations for the sounding of the atmosphere that make balloons launches usually at 0 UT but also at 12 UT in case of adverse weather. Data from all stations over the world 
can be retrieved from the University of Wyoming website, Department of Atmospheric Science\footnote{http://weather.uwyo.edu/upperair/sounding.html}. In our case, data from the weather stations 16080 
LIML (Milano), 16045 LIPI (Rivolto) and 16144 San Pietro Capofiume were taken. All the weather data from these stations were taken at 0 UT of May 31, 2017, about 3 hours after the fireball event. 
The nearest weather station to the terminal point of the luminous path was San Pietro Capofiume ($44^\circ 39^\prime 13.63^{\prime\prime}$ N; $11^\circ 37^\prime 22.28^{\prime\prime}$ E), about 100 km 
away. To compute the dark flight phase we simply take these last atmospheric data without the use of an atmospheric model to propagate it in space and time. Later in the text, we will estimate how a change in the wind regime can influence the strewn field center.\\
The motion of the residual meteoroid, starting from the observed terminal point of the luminous path, can be described using Newton's Resistance law as in Ceplecha \cite{Ceplecha1987}, because the meteoroid motion takes place in a turbulent regime, i.e. a motion characterized by high Reynolds number (see below), and the gravity force law. In gas dynamics physics the full vector equation of the meteoroid motion during dark flight is as follows:

\begin{equation}
    \vec a = -\Gamma\rho_a v \frac{A}{m}\vec v + \vec g
	\label{eq:newton_drag}
\end{equation}

with
\begin{equation}
    \vec v = \vec v_{c} - \vec V
	\label{eq:newton_vel}
\end{equation}

In the previous equations the symbols have the meaning listed in Table~\ref{tab:symbol2_table}.

\begin{table}
\centering
\caption{Meaning of the symbols for Eq.~(\ref{eq:newton_drag}) and (\ref{eq:newton_vel}).}
\label{tab:symbol2_table}
\begin{tabular}{ll}
\hline
Symbol & Quantity \\
\hline
$\Gamma$ & Aerodynamic drag coefficient\\
$\rho_a$ & Fluid density\\
$V$ & Fluid speed (in our case wind speed)\\
$v$ & Body speed with respect to the fluid\\
$v_c$ & Meteoroid speed with respect to the ground\\
$A$ & Meteoroid cross section after the ablation phase\\
$m$ & Meteoroid residual mass after the ablation phase\\
$g$ & Standard acceleration due to gravity\\
\hline
\end{tabular}
\end{table}

Making the substitution:

\begin{equation}
    \vec a = \frac{d\vec v}{dt} = \frac{d\vec v}{dh} \frac{dh}{dt} = \frac{d\vec v}{dh} v_{h}
	\label{eq:substitution}
\end{equation}

The Eq.~(\ref{eq:newton_drag}) takes the form:

\begin{equation}
    \frac{d\vec v}{dh} = -\frac{1}{v_h}\Gamma\rho_a v \frac{A}{m}\vec v + \frac{\vec g}{v_h} = -\frac{\Gamma\rho_a v}{D_{fin}v_h} \vec v + \frac{\vec g}{v_h}
	\label{eq:newton_drag2}
\end{equation}

To apply Eq.~(\ref{eq:newton_drag2}) in the real world it's assumed that ablation suddenly stops after the last observation was made. This may not be strictly true as the final point of the fireball trajectory could be just due to the observation range or to the sensitivity of the sensor that prevents from seeing the full fireball trajectory. Considering that the Rovigo station was very close to the terminal point of the fireball's trajectory (about 36 km), this effect is supposed to be not very important here.\\

\begin{figure}
\begin{center}
\resizebox{0.60\textwidth}{!}{\includegraphics{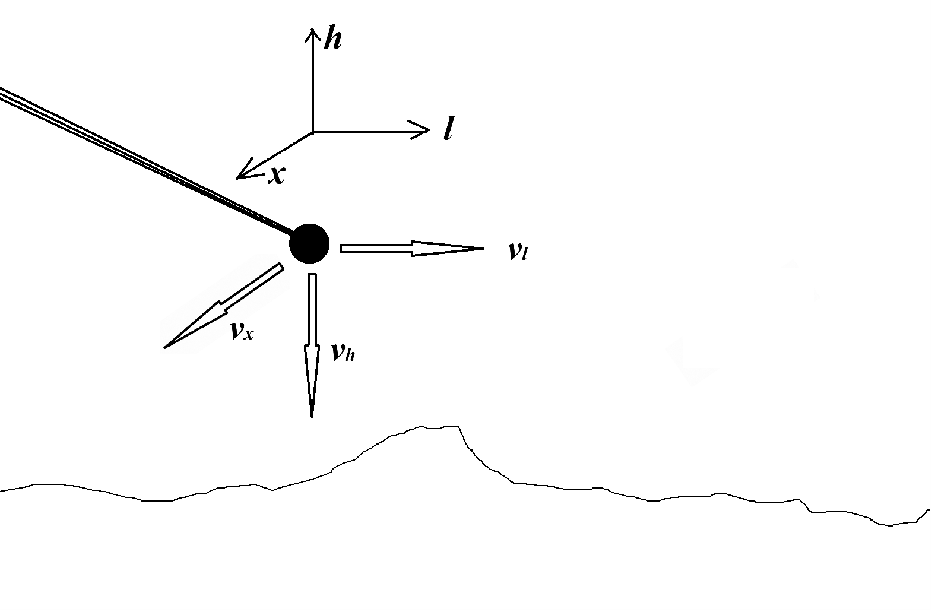}}
\end{center}
\caption{The fireball reference system ($l$, $h$, $x$) for dark flight phase. The black dot is the terminal point of the luminous path. The component velocity $v_l$ is parallel to the fireball 
motion direction, $v_x$ is along the orthogonal direction and $v_h$ is toward the bottom (so $v_h < 0$ always).}
\label{fig:Fireball_reference_system_dark_flight}
\end{figure}

\begin{figure}
\begin{center}
\resizebox{0.60\textwidth}{!}{\includegraphics{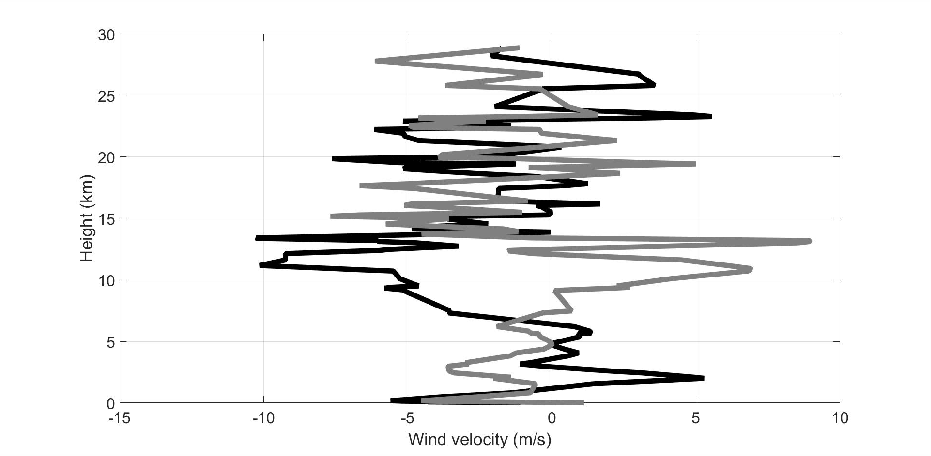}}
\end{center}
\caption{Wind speed vs. height in the meteoroid reference system from San Pietro Capofiume. \textbf{Black line}: wind along meteoroid motion direction ($l$ axis). Wind speed values greater than zero are against meteoroid motion. \textbf{Gray line}: right orthogonal direction to the meteoroid motion ($l_x$ axis). Wind speed values greater than zero are against the positive direction of $v_x$.}
\label{fig:Wind_velocity}
\end{figure}

Ceplecha also takes into account Coriolis-force, even if it is a small contribution and it is not shown in the previous equations \cite{Ceplecha1987}. In our numerical computations we also include the Earth's rotation. The reference system of the previous motion equations is shown in Fig.~\ref{fig:Fireball_reference_system_dark_flight}. The origin of this reference system is in the terminal point of the fireball path. Numerically integrating this differential equation with the winds values shown in Fig.~\ref{fig:Wind_velocity} we directly obtain meteoroid velocity vs. height above the ground. 

The ratio $m/A = D_{fin}$ and the $\Gamma$ value are given by the dynamical model of the previous section. The value of the aerodynamic drag coefficient $\Gamma$ depends both on the unknown final form 
of the meteoroid after ablation, on the Reynolds number and on the Mach number, i.e. the ratio between the meteoroid speed and the sound speed at the same height above ground. Assuming, for the residual 
meteoroid, a diameter $d$ of about 0.1 m (Table~\ref{tab:dynamic_table}), and taking into account air density and temperature in the terminal point of the fireball, a Reynolds number 
$R_e=\rho vd/\mu \approx 10^6$ can be estimated ($\mu$ is the dynamic viscosity of the fluid). That is, we are in a turbulent regime and this justifies resort to Newton's Resistance law. The same 
Reynold number still holds when the residual meteoroid touches the ground, because the decrease in speed is roughly compensated by the increase in air density (see  Table~\ref{tab:strewn_field_table}). 
In general - with Mach number between 8 and 20 - for a spherical body the drag coefficient $\Gamma$ decreases with the increase of the Reynolds number  towards an asymptotic value near 0.3-0.4 for 
$R_e > 10^4$ \cite{Bailey1971}. The value of the drag coefficient $\Gamma$ is independent of the size, the crucial parameter being the body shape. In our case the residual meteoroid will not be a 
perfect sphere so it is reasonable to expect that the $\Gamma$ values are a bit higher towards high Mach and Reynolds numbers.\\
For the asymptotic value, i.e. toward very high Mach numbers (Reynolds number is always high) the value $\Gamma = 0.58$, that we fix in the meteoroid dynamic model, appears reasonable. For low Mach 
numbers instead, i.e. equal or less than 4, we adopt the Ceplecha's values \cite{Ceplecha1987}: $\Gamma (4) = 0.58$, $\Gamma (3) = 0.62$, $\Gamma (2) = 0.63$, $\Gamma (1) = 0.50$, $\Gamma (0.8) = 0.44$,
$\Gamma (0.6) = 0.39$, $\Gamma (0.4) = 0.35$ and $\Gamma (0.2) = 0.33$. Finally, the horizontal distance along the $l$ axis, between the terminal point projected on the ground and the impact point is 
given by:

\begin{equation}
    L = \int_{0}^{T}{v_l dt} = \int_{h_T}^{h_S}{v_l\frac{dh}{v_h}} \Rightarrow L = \int_{h_S}^{h_T}{v_l\frac{dh}{v_h}} ~\textrm{ with } v_h > 0
	\label{eq:dark_flight}
\end{equation}

\begin{figure}
\begin{center}
\resizebox{0.60\textwidth}{!}{\includegraphics{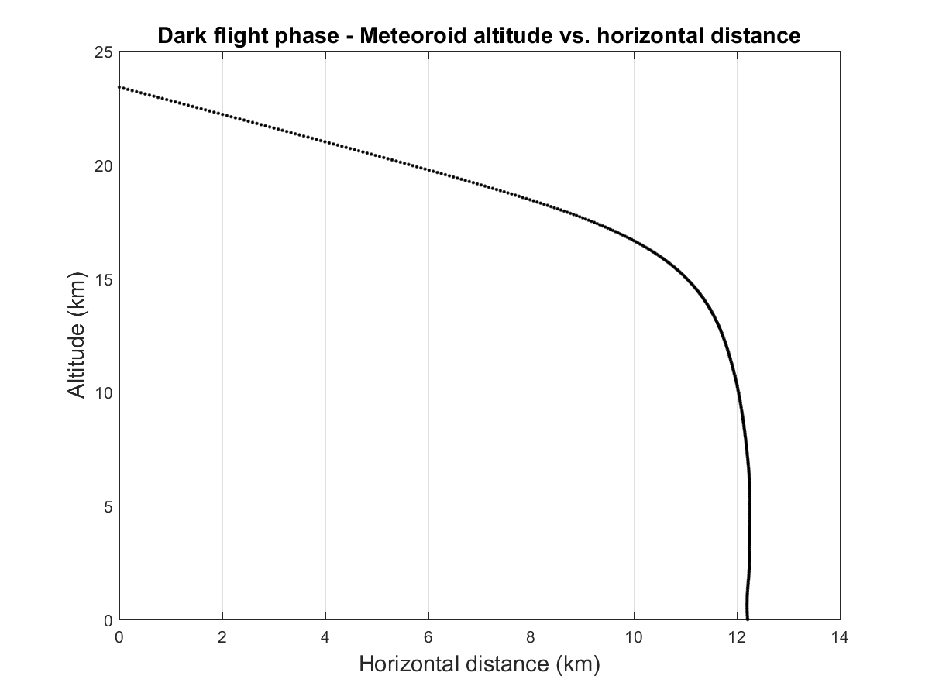}}
\end{center}
\caption{Parallel view of the residual meteoroid height vs. horizontal distance along the $l$ axis starting from the terminal point. Notice the small deformations at the end of the vertical section 
of the trajectory, due to the wind.}
\label{fig:Dark_flight_Meteoroid_altitude_vs_horizontal_distance}
\end{figure}

\begin{figure}
\begin{center}
\resizebox{0.60\textwidth}{!}{\includegraphics{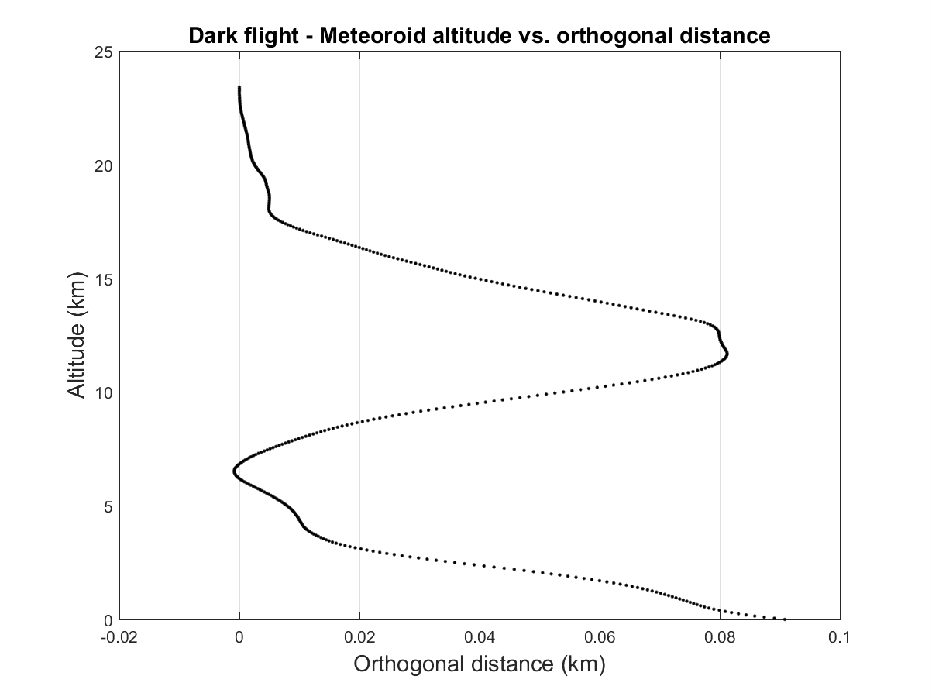}}
\end{center}
\caption{Front view of the residual meteoroid trajectory. Meteoroid height vs. orthogonal distance $l_x$. The trajectory oscillations are in phase with the winds directions but the orthogonal movements 
are a few tens of meters only.}
\label{fig:Dark_flight_Meteoroid_altitude_vs_orthogonal_distance}
\end{figure}

A similar equation holds for $L_x$, the orthogonal displacement \cite{Ceplecha1987}. These distances, $L$ and $L_x$, were computed with numerical integration of the motion equations assuming, as 
starting conditions, the position, velocity and acceleration given in Table~\ref{tab:dynamic_table} with the dynamical model computed at the terminal point of the fireball 
(see Fig.~\ref{fig:Dark_flight_Meteoroid_altitude_vs_horizontal_distance} and Fig.~\ref{fig:Dark_flight_Meteoroid_altitude_vs_orthogonal_distance}).

\begin{figure}
\begin{center}
\resizebox{0.60\textwidth}{!}{\includegraphics{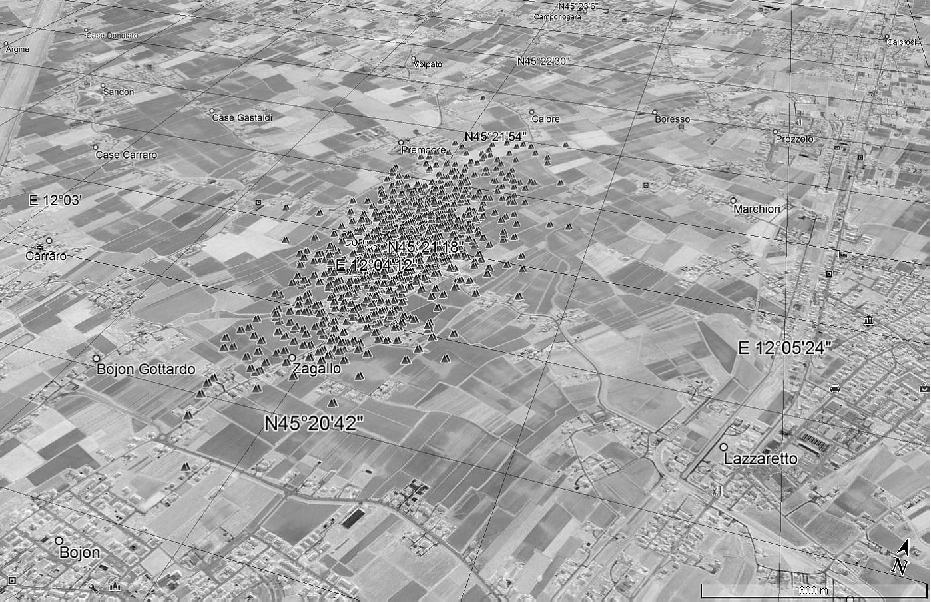}}
\end{center}
\caption{A Google Earth maps showing the possible impact points on the ground obtained from the Monte Carlo simulation. The strewn field dimension is about $1.7 \times 0.6$ km, about north of the town of Bojon.}
\label{fig:IT20170530_impact_points}
\end{figure}

From our computation we found that the Mach number in the terminal point of the fireball phase is about 10. Mach numbers falls below 4 only in the last 20 km above the ground. To demarcate the 
probable impact zone we have chosen three different values for the $m/A$ ratio in the terminal point (see Table~\ref{tab:strewn_field_table}), according with the uncertainty given in Table~\ref{tab:dynamic_table}, and seen how the different impact points are distributed on the ground. According to the $m/A$ values, the distance of the impact point from the projection on the ground of the terminal 
point varies from 11.9 to 12.5 km with a difference of about 0.6 km. The impact velocity with the ground is around 76 m/s, i.e. about 274 km/h. The uncertainty about velocity and height in the 
terminal point have a minor influence over the impact point. We have delimited the full strewn field using a Monte Carlo simulation, i.e. creating 1000 virtual meteoroids with parameters compatible 
with the observations in the terminal point and computing for each of them the point of fall. The full strewn field has an extension of about $1.7 \times 0.6$ km 
(see Fig.~\ref{fig:IT20170530_impact_points}).

\begin{table}
\centering
\caption{Data regarding the impact points with different $m/A$ final values compatible with the uncertainty given in 
Table~\ref{tab:dynamic_table}.}
\label{tab:strewn_field_table}

\begin{tabular}{lccc}
\hline
Quantity & \\
\hline
Final $m/A$ (kg/$\textrm{m}^2$)   & 210	      & 220     &	230 \\
Lat. N impact point ($^\circ$)	  &	45.3522	  & 45.3546	&   45.3570\\
Long. E impact point ($^\circ$)   &	12.0705	  & 12.0710	&   12.0715\\
$L$ (km)	                      & 11.9	  & 12.2	&   12.5\\
$v_{impact}$ (m/s)	              & 74	      & 76      &	78\\
\hline
\end{tabular}
\end{table}

Considering that we used weather data 100 km away in space and 3 hours in time from the place and instant of the fireball fall, the weak point of these results about the strewn field is that the assumed wind regime probably is not similar to that really present during the fall. So we have made a rough estimate of how important is to know the exact atmospheric state to compute the strewn field. We recompute the dark flight using the data from the weather stations 16080 (Milano, 250 km away), 16045 (Rivolto, 100 km away) and 16144 (Capofiume), both for 0 UT of May 30 and 31. The result is that the six nominal impact points are very close, with a standard deviation of about 0.5 km in latitude and 0.4 km in longitude. So it can be expected that the change in wind speed have shifted the nominal impact point by about 0.6 km in any direction. For future interesting fireballs, it will be desirable to use atmospheric models to obtain the wind regime and the state of the atmosphere for the desired place and time in order to reduce strewn field uncertainty.

\section{In search for meteorites}
\label{sec:search} 

After numerical computation of the possible impact points on the ground, we looked for meteorites. Immediately after the fall the strewn field was wider than indicated in this paper, because the speed at the end point was estimated with simple kinematic considerations. Only after introducing the dynamic model for the meteoroid was the search area better delimited.\\ 
Public appeals have been made to the population of the areas involved, including on the PRISMA website\footnote{http://www.prisma.inaf.it/index.php/2017/06/27/bolide-del-30-maggio-era-un-mini-asteroide-segnalateci-eventuali-sassi-strani-o-anomali/} in several newspapers as well as on social media. Following these appeals, over 10 suspected meteorites have been collected by local inhabitants. The samples have been all identified as common ground stones. We also did directly search the predicted strewn field starting a few days after the fall until the early July 2017, and a second search was done in April 2018. A great contribution for meteorites search came from ``Meteoriti Italia'', a group of amateur meteorite enthusiasts who like to support researchers, contributed greatly by assisting on the meteorite search trips. Unfortunately the area where the ``on field research'' 
took place is densely populated and settled with  villages, streets and water channels. Moreover, it is also a place of an intensive agricultural activity. There were several crops in progress, 
including wheat fields, that could not be accessed until after harvest. This ``difficult territory'' has hindered searches and no meteorites were found. The only collected objects, at first sight 
similar to a meteorite, were some rounded fragments of black volcanic glass. Probably the glass originated from the ancient volcanos that gave origin to a hills site called ``Colli Euganei'' about 
30 million years ago \cite{Piccoli1981}, and located about 30 km away from the computed impact points. However, we do not rule out the possibility to find a meteorite in a near future with more 
thorough searches.

\section{The meteoroid heliocentric orbit and the search for a progenitor body}
\label{sec:orbit} 

\begin{figure}
\begin{center}
\resizebox{0.60\textwidth}{!}{\includegraphics{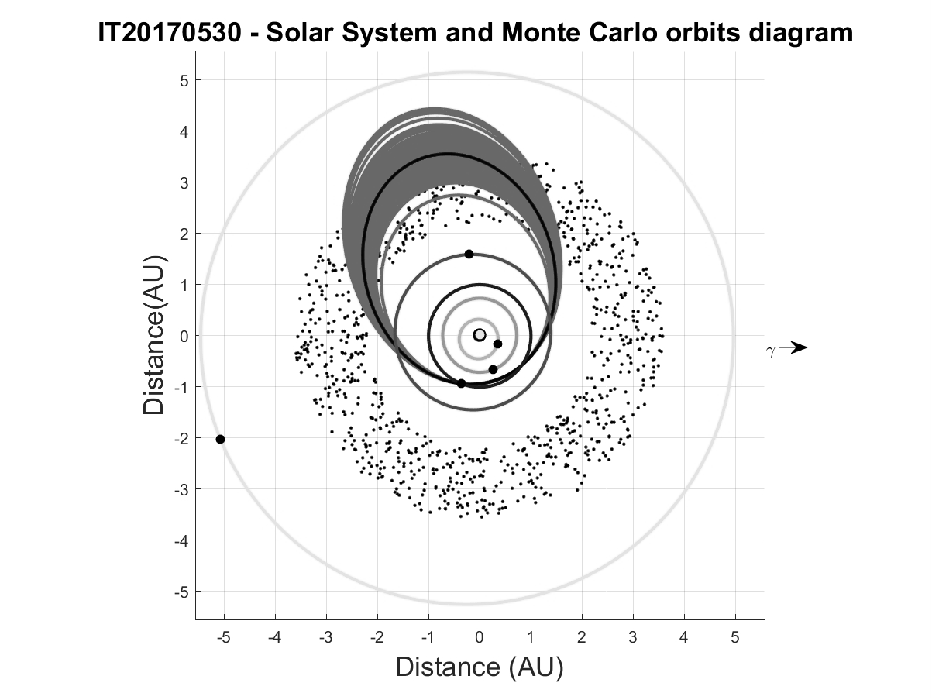}}
\end{center}
\caption{The nominal heliocentric orbit for the progenitor meteoroid of the fireball IT20170530 as seen from the ecliptic north pole. The dots symbolically represent the Main Belt. The position 
of the planets on their orbits is that at the time of the fireball. The fireball clones orbit are indicated in gray color.}
\label{fig:Solar_System_and_Monte_Carlo_orbits_diagram}
\end{figure}

Knowing the heliocentric velocity vector of the progenitor meteoroid and the Earth's vector position at the time of the meteoroid fall, it is possible to compute the heliocentric orbital elements 
\cite{Sterne1960, Ceplecha1987}. As to our case, it is interesting to note that a comparison between Ceplecha analytical orbit determination method and numerical integration yields consistent results 
\cite{Clark2011}.\\  
Of course uncertainty about the heliocentric speed, both in length and direction, also makes the orbital elements uncertain (see Table~\ref{tab:orbit_table} and 
Fig.~\ref{fig:Solar_System_and_Monte_Carlo_orbits_diagram}). In order to estimate the uncertainty of the orbital elements, a Monte Carlo approach with 100 clones was performed. The computed 
orbital elements indicate that the meteoroid was an Apollo-type object, with an aphelion near the outer Main Belt and with low inclination above the Ecliptic plane. With the data from 
Table~\ref{tab:orbit_table}, the heliocentric distance of the ascending node was about 1.022 UA, whereas the descending node was near 2.81 AU. Incidentally, we note that the heliocentric distance 
of the ascending node is consistent with that of the Earth on May 30th, 2017 i.e. 1.014 AU.

In order to identify a possible parent body among the known NEAs, we use the $D_N$ criterion introduced by \cite{Valsecchi1999} for meteoroid stream identification:
\begin{equation}
    D_N = \sqrt{{\left( U-U_0 \right)}^2+{\left( \cos\theta-\cos\theta_0 \right)}^2+{\left( 2\sin\frac{\phi-\phi_0}{2}\right)}^2+{\left( 2\sin\frac{\lambda-\lambda_\oplus}{2}\right)}^2}
    \label{eq:valsecchi}
\end{equation}
At variance with most other criteria, based on the heliocentric orbital elements, this criterion uses geocentric quantities and two of the quantities that are used in $D_N$ (i.e. $U$ and $cos\theta$), have been shown to be nearly invariant under the secular perturbation. Many factors influence the dynamical evolution of a meteoroid, and some of them result from forces other than gravitation, especially for meteoroids of very small size. However, over not too long time-scales, and in the absence of planetary close encounters, we can assume that only planetary secular perturbations affect meteoroid orbits. For this reason we consider the $D_N$ criterion useful for finding a progenitor. We refer the reader to \cite{Valsecchi1999} for the details, and to \cite{Galligan2001} and \cite{Moorhead2016} for comparisons with other criteria.

\begin{table}
\centering
\caption{Data about the meteoroid heliocentric orbit. The standard deviations are obtained with a Monte Carlo computation over 100 clones. The longitude of the ascending node has very low 
uncertainty because the value is only determined by the time of the fireball fall.}
\label{tab:orbit_table}
\begin{tabular}{lc}
\hline
Quantity & Numerical value\\
\hline
Semi major axis (AU)                       & $2.3 \pm 0.2$ \\
Eccentricity	                           & $0.59 \pm 0.03$\\
Orbital Period (years)	                   & $3.4 \pm 0.4$\\
Orbit inclination ($^\circ$)               & $4.2 \pm 0.1$\\
Longitude of the ascending node ($^\circ$) & $249.4002 \pm 0.0001$\\
Argument of Perihelion ($^\circ$)          & $37.7 \pm 0.1$\\
Perihelion passage (JD)	                   & $2456672.9 \pm 138$\\
Perihelion distance (AU)                   & $0.94 \pm 0.01$\\
Aphelion distance (AU)                     & $3.6 \pm 0.3$\\
\hline
\end{tabular}
\end{table}

\begin{table*}
\centering
\caption{NEAs with geocentric parameters $U$, $\theta$, $\phi$ and $\lambda$ close to the IT20170530 values. The uncertainty on the asteroids elements are one or more orders of magnitude lower than that of the meteoroid.}
\label{tab:NEAs_parameters_table}

\begin{tabular}{llccccc}
\hline
 & NEA & $U$ & $\theta (^\circ)$ & $\phi (^\circ)$ & $\lambda (^\circ)$ & $D_N$\\
\hline
         & 2011 UR$_{63}$  & 0.40 & 60$^\circ\!\!$.0 & 285$^\circ\!\!$.3 & 251$^\circ\!\!$.3 & 0.067 \\
         & 2017 WD         & 0.35 & 55$^\circ\!\!$.9 & 285$^\circ\!\!$.5 & 254$^\circ\!\!$.7 & 0.072 \\
         & 2018 VL$_3$     & 0.39 & 54$^\circ\!\!$.7 & 280$^\circ\!\!$.6 & 255$^\circ\!\!$.3 & 0.101 \\
         & 2008 TQ$_{26}$  & 0.34 & 61$^\circ\!\!$.9 & 288$^\circ\!\!$.3 & 245$^\circ\!\!$.3 & 0.112 \\
         & 2017 WO$_{13}$  & 0.39 & 59$^\circ\!\!$.8 & 278$^\circ\!\!$.3 & 248$^\circ\!\!$.1 & 0.113 \\
         & 2018 WT$_1$     & 0.32 & 55$^\circ\!\!$.1 & 291$^\circ\!\!$.3 & 242$^\circ\!\!$.1 & 0.126 \\
         & 2019 EU         & 0.35 & 55$^\circ\!\!$.0 & 289$^\circ\!\!$.5 & 259$^\circ\!\!$.1 & 0.132 \\
         & 2017 KW$_4$     & 0.42 & 62$^\circ\!\!$.1 & 292$^\circ\!\!$.4 & 246$^\circ\!\!$.5 & 0.132 \\
         & 2017 PL$_{26}$  & 0.30 & 54$^\circ\!\!$.9 & 280$^\circ\!\!$.2 & 255$^\circ\!\!$.3 & 0.134 \\
         & 2017 KR$_{27}$  & 0.41 & 60$^\circ\!\!$.5 & 290$^\circ\!\!$.6 & 241$^\circ\!\!$.0 & 0.139 \\
         & 2011 UD$_{115}$ & 0.32 & 60$^\circ\!\!$.7 & 293$^\circ\!\!$.8 & 245$^\circ\!\!$.8 & 0.141 \\
         & 2012 VT$_{76}$  & 0.39 & 60$^\circ\!\!$.8 & 277$^\circ\!\!$.5 & 254$^\circ\!\!$.6 & 0.143 \\
         & 2011 PO$_1$     & 0.42 & 60$^\circ\!\!$.3 & 282$^\circ\!\!$.7 & 258$^\circ\!\!$.8 & 0.146 \\
         & 2005 XO$_4$     & 0.37 & 54$^\circ\!\!$.3 & 277$^\circ\!\!$.5 & 257$^\circ\!\!$.8 & 0.149 \\
(523685) & 2014 DN$_{112}$ & 0.32 & 55$^\circ\!\!$.7 & 289$^\circ\!\!$.5 & 239$^\circ\!\!$.0 & 0.149 \\
         & 1997 UA$_{11}$  & 0.40 & 59$^\circ\!\!$.4 & 281$^\circ\!\!$.4 & 239$^\circ\!\!$.1 & 0.150 \\
\hline
         &                 & $U$  & $\theta$          & $\phi$            & $\lambda_\oplus$ \\
\hline
\multicolumn{2}{c}{IT20170530} 
                           & 0.38 &  56$^\circ\!\!$.0 & 286$^\circ\!\!$.0 & 249$^\circ\!\!$.4 \\
\hline
\end{tabular}
\end{table*}

\begin{figure}
\begin{center}
\resizebox{0.50\textwidth}{!}{\includegraphics{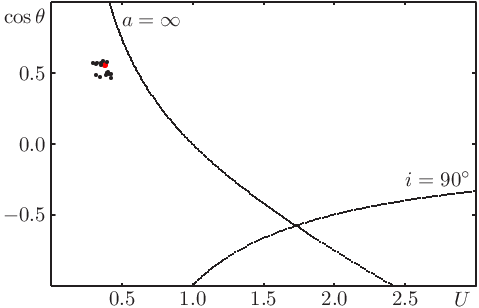}}
\end{center}
\caption{The black dots shows the NEAs with $D_N<0.15$ in the plane $U$-$\cos\theta$, the big red dot shows IT20170530. The region below the $a=\infty$ line contains orbits bound to the Sun, and the region on the left of the $i=90^\circ$ line contains prograde orbits.}
\label{fig:U_cos_theta_IT20170530}
\end{figure}

For the NEAs, the relevant quantities are conveniently tabulated by NEODyS\footnote{https://newton.spacedys.com/~neodys2/propneo/encounter.cond}; for IT20170530 we have 
$U_0 = 0.38 \pm 0.02$, $\theta_0 = 56.0^{\circ} \pm 0.2^{\circ}$, $\phi_0 = 286^{\circ} \pm 0.7^{\circ}$ and $\lambda_\oplus=249.4^{\circ}$ (the longitude of the Earth at the time of fall).

Table~\ref{tab:NEAs_parameters_table} reports the NEAs characterized by $D_N<0.15$ with respect to IT20170530; the same NEAs are shown in the $U$-$\cos\theta$ plane in Fig.~\ref{fig:U_cos_theta_IT20170530}. Practically all of these NEAs are small to very small objects, characterized by very low values of their MOID (Minimum Orbit Intersection Distance); the exception is (523685)~2014 DN$_{112}$, a numbered object characterized by $H=20.0$.  However, none of the orbits of the NEAs in the Table is particularly close to the orbit of the fireball.
However, Fig.~\ref{fig:U_cos_theta_IT20170530} and Table~\ref{tab:NEAs_parameters_table} show that the meteoroid was in a region populated by small NEAs, which suggests a possible asteroidal origin. We plan to continue to scan the NEAs database to see if new asteroids, with lower $D_N$ values, will be discovered.

\section{Conclusions}
\label{sec:end} 

We have presented the main results about the fireball IT20170530, observed by PRISMA, IMTN and CMN stations on May 30th, 2017 at about 21h 09m 17s UTC. Unfortunately only data from 
the Rovigo station appear to be the most complete and usable, which  represented a significant shortcoming in  the analysis. However, according to our results, the progenitor meteoroid 
entered the atmosphere at a speed $v_\infty = 15.9 \pm 0.3$ km/s, with an estimated starting mass/section ratio $D_\infty = 234 \pm 15$ kg/$\textrm{m}^2$. If the body was a spherical chondrite with mean drag coefficient $\Gamma$ = 0.58, we estimated a guess starting diameter of about 0.1 m and a mass of about 1.8 kg. Thanks to the low relative speed with the Earth the ablation was slow and the dynamic model indicates that a residual meteoroid is possible because $D_{fin} > 0$.\\
The fireball path extinct at a terminal height $H_t = 23.3 \pm 0.2$  km (Lat. $45.246^\circ \pm 0.002^\circ$ N; Long. $12.046^\circ \pm 0.002^\circ$ E), between the Italian cities of Venice and Padua. The dark flight phase led the residual meteoroid, of about 0.09 m diameter and mass 1.6 kg (guess values), to fall about 11.9-12.5 km beyond the trajectory terminal point. The effect of the winds and wind variation on the fall was several hundred of meters at most. Also important is the effect of the final mass/cross section ratio uncertainty that has led us to delimit a minimum strewn field of about $1.7 \times 0.6$ km. In this and a larger area we searched unsuccessfully for meteorites. The progenitor meteoroid heliocentric orbit indicates that the body came from the outer Main Belt of asteroids but it is uncertain because the speed values come from the Rovigo station only. The search for a specific progenitor body among the known NEAs has not given good candidates, but we plan to continue to scan the NEAs database to see if new asteroids, with lower $D_N$ values, will be discovered.\\

The physical analysis of the fireballs set out in this paper will serve as a reference for future events. Thanks to the great expansion of PRISMA network in Italy, we hope to have interesting events whose data come only from PRISMA stations, in order to have maximum data homogeneity. In the case of IT20170530, having non-homogeneous data certainly was not good, for example as regards the measure of speed versus time. The lack of usable photometric data concerns only this specific case, we hope to be able to obtain the lightcurves of the fireballs from the PRISMA cameras far enough away that they are not saturated. \\
The implementation of an automatic pipeline for PRISMA is in progress. It would be necessary to have a real-time alert system which, depending on the fireball final height, warns if the fireball extinguishes below 25-30 km from ground. These are the events where meteorites are most likely. An automatic alert system would allow us to arrive as soon as possible to look for the meteorite in the strewn field, minimizing terrestrial contamination. In this case no meteorites were found but it happened with the very recent fireball IT20200101 at 18:26:54 UT. This historic event will be the subject of a next paper.

\section*{Acknowledgements}

The authors wish to thank all the owners and managers of the PRISMA stations that with their support make possible the study of fireballs and the search for meteorites. The complete list of people,
associations and institutions (both public and private) is available on the PRISMA website: http://www.prisma.inaf.it. PRISMA was partially funded by a 2016 ``Research and Education'' grant from 
Fondazione CRT. Many thanks to J. Belasl and D. \v Segonl of the Croatian Meteor Network for giving us their data about IT20170530. Also many thanks to U. Repetti, chairman of ``Meteoriti Italia'', 
for the work about meteorite search on the strewn field.\\

%
%

\end{document}